\documentclass[twocolumn]{aastex62}

\usepackage{xspace}

\newcommand{\Ha}{H$\alpha$\xspace}

\newcommand{\kube}{{\sc kubeviz}\xspace}
\newcommand{\sinopsis}{{\sc sinopsis}\xspace}

\defcitealias{Poggianti2017a}{Paper I}
\defcitealias{Bellhouse2017}{Paper II}
\defcitealias{Fritz2017}{Paper III}
\defcitealias{Gullieuszik2017}{Paper IV}
\defcitealias{Moretti2018}{Paper V}
\defcitealias{Vulcani2017c}{Paper VIII}
\defcitealias{Vulcani2018}{Paper VII}
\defcitealias{Jaffe2018}{Paper IX}

\defcitealias{Poggianti2016}{P16}

\graphicspath{{./}{figures/}}

\received{August 30, 2018}
\revised{September 26, 2018}
\accepted{October 4, 2018}
\submitjournal{ApJL}

\shorttitle{SFR-Mass relation in the GASP sample}
\shortauthors{Vulcani et al.}

\begin{document}

\title{Enhanced star formation in both disks and { ram pressure stripped} tails of GASP jellyfish galaxies}

\correspondingauthor{Benedetta Vulcani}
\email{benedetta.vulcani@inaf.it}

\author[0000-0003-0980-1499]{Benedetta Vulcani}
\affil{INAF- Osservatorio astronomico di Padova, Vicolo Osservatorio 5, IT-35122 Padova, Italy}
\author{Bianca M. Poggianti}
\affil{INAF- Osservatorio astronomico di Padova, Vicolo Osservatorio 5, IT-35122 Padova, Italy}
\author{Marco Gullieuszik}
\affil{INAF- Osservatorio astronomico di Padova, Vicolo Osservatorio 5, IT-35122 Padova, Italy}
\author{Alessia Moretti}
\affil{INAF- Osservatorio astronomico di Padova, Vicolo Osservatorio 5, IT-35122 Padova, Italy}
\author{Stephanie Tonnesen}
\affil{Center for Computational Astrophysics, Flatiron Institute, 162 5th Ave, New York, NY 10010, USA}
\author{Yara L. Jaff\'e}
\affiliation{Instituto de F\'isica y Astronom\'ia, Universidad de Valpara\'iso, Avda. Gran Breta\~na 1111 Valpara\'iso, Chile}
\author{Jacopo Fritz}
\affiliation{Instituto de Radioastronom\'ia y Astrof\'isica,
UNAM, Campus Morelia, A.P. 3-72, C.P. 58089, Mexico}
\author{Giovanni Fasano}
\affil{INAF- Osservatorio astronomico di Padova, Vicolo Osservatorio 5, IT-35122 Padova, Italy}
\author{Daniela Bettoni}
\affil{INAF- Osservatorio astronomico di Padova, Vicolo Osservatorio 5, IT-35122 Padova, Italy}

\begin{abstract}
Exploiting the data from the GAs Stripping Phenomena in galaxies with MUSE (GASP) program, we compare the integrated Star Formation Rate- Mass relation (SFR-M$_\ast$) relation of 42 cluster galaxies undergoing ram pressure stripping (``stripping galaxies'') to that of 32 field and cluster undisturbed galaxies. Theoretical predictions have so far led to contradictory conclusions about whether ram pressure can enhance the star formation in the  gas disks and tails or not and until now a statistically significant observed sample of stripping galaxies was lacking. We find that stripping galaxies occupy the upper envelope of the control sample SFR-M$_\ast$ relation, showing a systematic  enhancement of the SFR at any given mass. The star formation enhancement occurs in the disk (0.2 dex), and additional star formation takes place in the tails. Our results suggest that strong ram pressure stripping events can  moderately enhance the star formation also in the  disk prior to gas removal.

\end{abstract}

\section{Introduction} \label{sec:intro}

Star forming galaxies falling onto clusters can lose their gas via ram pressure stripping (RPS) due to their motion through the intracluster medium \citep[ICM;][]{Gunn1972}.
As the gas is lost, the star formation (SF) gets quenched and galaxies eventually become passive. 

Observationally, there is increasing evidence for galaxies observed at different stages of  stripping. The most spectacular cases, at the peak of the stripping, are the so called jellyfish galaxies, which show tails with ionised gas and bright blue knots downstream of the disks, indicating substantial SF in their tails, and asymmetric disks of young stars \citep[e.g.][]{Cortese2007, Smith2010, Fumagalli2014, Fossati2016, Consolandi2017, Poggianti2017a}. 

{Few observational works, based on individual objects, have shown that the RP enhances the SF before quenching it \citep{Crowl2006, Merluzzi2013, Kenney2014}, but the effect has  never been quantified in a statistically significant sample. On the other hand, \cite{Crowl2008}, analysing 10 Virgo galaxies that underwent RPS, found that  in general, there is, at most, a modest starburst prior to quenching. }

\cite{Poggianti2016}  assembled a catalog of local gas stripping candidates, based on B-band images,  and found that galaxies showing signs of  stripping are preferentially located above the typical Star Formation Rate - Mass (SFR-M$_\ast$) relation, indicating a SFR excess with respect to normal star forming galaxies of the same mass. However, their work was based on integrated quantities and the derivation of SFRs and masses were performed using single fiber spectroscopy, which is affected by aperture losses  that need to be corrected adopting some assumptions on the mass-to-light ratio gradient. In addition, their SFR measurements 
did not take into account that the presence of active galactic nucleus (AGN) and old (post-AGB) stellar population could alter their estimates \citep{YanBlanton2012, Belfiore2016}. 
As the fraction of AGN is very high among stripping candidates (\citealt{Poggianti2017b}, Radovich et al. in prep.) an incorrect treatment of the AGN can affect the SFR-M$_\ast$ relation. 

Using simulations, several groups have been studying the effect of RPS on individual galaxies, often focusing on its impact on the galaxy SFR \citep[e.g.,][]{FujitaNagashima1999, Quilis2000, RoedigerHensler2005, RoedigerBruggen2006, Tonnesen2011,  TonnesenBryan2012}. Results have been mixed; for example,  \cite{Kronberger2008, Kapferer2009} found that RPS enhances the overall SFR  by up to a factor of 3-10 with respect to an isolated galaxy. \cite{Kronberger2008} found that new stars are mainly formed in the central parts of the disk but a significant fraction forms in the wake of the galaxy, while \cite{Kapferer2009} found a shift in the SF from the disk to the wake, with net SFR suppression in the disk. The simulations of \cite{TonnesenBryan2012}  did not find a significant enhancement of SF in the remaining gas disk, and only low levels of SF in the stripped tails. \cite{Roediger2014} showed that SF enhancements take place only in regions of sufficiently low initial interstellar medium pressure, which will be stripped soon afterwards. \cite{Bekki2014, Steinhauser2016} found that SF enhancement can occur in RP stripped galaxies, although it depends strongly on the satellite mass, orbit, and inclination angle.

\cite{Troncoso2016}, using thee EAGLE simulation, showed that the SF enhancement occurs only in the half of the galaxy that is facing the cluster center during its infall. Fritz et al. (in prep.)  found  observational evidence for this result.  Analysing the spatially resolved SFR of jellyfish galaxies in their initial stripping phase, they found a SFR enhancement on the leading side of the galaxy disk. 
Assuming that galaxies move towards the cluster center, this relative triggering corresponds to the RPS compression. 

The physical  origin of the enhancement is still not totally understood. 
In dynamically disturbed clusters extreme RPS events can be abundant \citep{VijayaraghavanRicker2013, Jaffe2016, McPartland2016} and the  SFR triggering seems to be higher in merging clusters \citep[but see][for opposite results]{FujitaNagashima1999}. In these environments, 
 galaxies can encounter higher velocity ICM head winds after being overrun by the merger shocks. The accompanying enhanced ICM pressure behind the shock could potentially boost the SF in these stripping galaxies \citep{BekkiCouch2010, Bekki2010, Roediger2014}. The higher the compression, the faster the SF quenching \citep{Bekki2010}. 
 
In this paper we analyse the disk SFR-M$_{\ast}$ relation of stripping and undisturbed galaxies, using IFU data covering up to several effective radii from the galaxy disk.  
Our sample is extracted from GASP\footnote{\url{http://web.oapd.inaf.it/gasp/index.html}} (GAs Stripping Phenomena in galaxies with MUSE),  an ESO Large programme granted 120 hours of observing time with the integral-field spectrograph MUSE that was completed in April 2018. GASP allows us to study galaxies in the local Universe in
various stages of RPS in clusters \citep{Jaffe2018}, from
pre-stripping (undisturbed galaxies), to initial
stripping, peak stripping \citep{Poggianti2017a, Bellhouse2017, Gullieuszik2017, Moretti2018}, and post-stripping \citep{Fritz2017}, passive and devoid of gas.

While many single jellyfish galaxies have been already studied in the literature \citep[e.g.][]{Cortese2007, Smith2010, Merluzzi2013, Fumagalli2014, Fossati2016, Consolandi2017}, GASP provides us with the unique possibility of looking for trends and perform comparisons in an homogeneous sample, reducing  possible biases. 
Apart from \cite{Poggianti2016}, no attempts to place them on the SFR-mass relation have been carried out so far.

We adopt a \cite{Chabrier2003} initial mass function (IMF) in the mass range 0.1-100 M$_{\odot}$. The cosmological constants assumed are $\Omega_m=0.3$, $\Omega_{\Lambda}=0.7$ and H$_0=70$ km s$^{-1}$ Mpc$^{-1}$. 

\section{The data sample} \label{sec:data}

The GASP targets are at redshift $0.04<z<0.1$, are located in
different environments (galaxy clusters, groups, filaments and
isolated) and span a wide range of galaxy stellar masses, from $10^9$
to $10^{11.5} M_{\odot}$. They were drawn from the \cite{Poggianti2016} catalog, based on 
the WINGS \citep{Fasano2006} and OMEGAWINGS \citep{Gullieuszik2015} cluster surveys and the Padova Millennium Galaxy and Group Catalogue \citep[PM2GC,][]{Calvi2011}. The GASP sample comprises 94 galaxies selected as stripping candidates (64 in clusters and 30 in groups, filaments or isolated), plus another 20 undisturbed galaxies in clusters or the field selected as a control sample.
A complete description of the survey strategy, data reduction and analysis procedures is presented in \citet[][Paper I]{Poggianti2017a}. 

In this work, we have selected cluster members with signs of initial (J0.5), moderate  (J1), {and extreme (J2) stripping, as well as  truncated disks (J3),} for a total of 42 galaxies. We disregarded all the uncertain cases. Details on the stripping stages (Jstage)  can be found in \cite{Jaffe2018}. We will call this sample `stripping sample'. 
{The subdivision into Jstages aims at distinguishing galaxies in different evolutionary phases and at different stages of gas stripping. J3 (9.5\% of the total)  have very little gas left in the disk and  are at a later stripping stage than the other galaxies. The J0.5 (9.5\%) and J1 (45\%) still have significant gas in the disk, and the J2 (36\%) display long tails of stripped material.}

We have then extracted from the cluster control sample and the field those galaxies that indeed are undisturbed and do not show any sign of environmental effects (RPS, tidal interaction, mergers, gas accretion...) on their spatially resolved SF distribution, for a total of 17 cluster members and  15 field galaxies.  
{While the control sample is relatively small, it ensures  control over the systematics that can arise from comparing inhomogeneous samples. Indeed, differences in, e.g., assumed cosmology,  IMF, luminosity-to-SFR  conversions, stellar population models, dust attenuation, and emission line contributions can lead to differences in derived stellar masses and SFRs as high as a factor of two-three (e.g. Speagle et al. 2014, Davies et al. 2016).}

\subsection{Observations and Data reduction}
Observations were carried out in service mode with the MUSE spectrograph mounted at the VLT. MUSE has 0.2$^{\prime\prime}\times$0.2$^{\prime\prime}$ pixels and covers a 1$^{\prime}\times$1$^{\prime}$ field-of-view. It covers the spectral range between 4800 and 9300 \AA{} sampled at 1.25 \AA{}/pixel with a spectral resolution FWHM=2.6\AA{}. Most of the targets were observed with one MUSE pointing, while for some of them two pointings were needed to cover the entire galaxy and the tail. On each pointing, 4$\times$675sec exposures were taken in clear, dark-time, $<1^{\prime\prime}$ seeing conditions. 

The data were reduced with the most recent available version of the MUSE pipeline\footnote{\url{http://www.eso.org/sci/software/pipelines/muse}}, as described in \citetalias{Poggianti2017a}.  All datacubes were then averaged filtered  in the spatial direction with a 5$\times$5 pixel kernel,  corresponding to our worst seeing conditions of 1$^{\prime\prime}$  = 0.7-1.3 kpc at the redshifts of the GASP  galaxies
\citepalias[see][for details]{Poggianti2017a}.

\subsection{Data analysis}
The methods used to analyse  the GASP datacubes are extensively presented in \citetalias{Poggianti2017a}. In brief, we corrected the reduced datacube for extinction due to our Galaxy and subtracted the stellar-only component of each spectrum
derived with our spectrophotometric code \sinopsis \citep{Fritz2017}.  
This tool also 
 provides for each MUSE spaxel many different quantities. Stellar masses, along with errors,  are the ones of interest in this work. 

We then derived emission line fluxes with associated errors using \kube \citep{Fossati2016}, an IDL public software. 
\Ha luminosities corrected both for stellar absorption and for dust
extinction were used to compute SFRs, adopting the \cite{Kennicutt1998a}'s relation: $\rm SFR (M_{\odot}
\, yr^{-1}) = 4.6 \times 10^{-42} L_{\rm H\alpha} (erg \, s^{-2})$. 
The extinction is estimated from the Balmer decrement
assuming a value $\rm H\alpha/H\beta = 2.86$ and the \cite{Cardelli1989} extinction law.\footnote{{ Note that even though we do not take into account the galaxy inclination angles in the computation of the extinction, both the control and the stripping sample have similar distributions and median values.}} As the formal errors obtained by \kube are negligible with respect to the uncertainties of the conversion factor from luminosities to SFR, we assume uncertainties on SFR to be 20\% of the values (Kennicutt et al. 2009).

We employed the standard diagnostic diagrams
[OIII]5007/$\rm H\beta$ vs [NII]6583/$\rm H\alpha$ to separate the regions powered by SF, Composite (SF+AGN), AGN and LINER emission. We adopted the division lines by \citet{Kauffmann2003}, \citet{Kewley2001},  and \citet{Sharp2010}. To compute SFRs, we considered only the spaxels whose ionised flux is powered by SF or  belong to the Composite region defined by \cite{Kauffmann2003}.

Both for stellar masses and for SFRs, we computed total integrated quantities by summing the values of all the spaxels belonging to each galaxy. To determine the galaxy disk, we use the definition of galaxy boundaries developed by Gullieuszik et al. (in prep.) and already exploited by Poggianti et al. (submitted). For each galaxy, these boundaries are computed from the map of the stellar continuum in the \Ha region and from the isophote with a surface brightness 1$\sigma$ above the average sky background level. Because of the (stellar and gaseous) emission from the stripped gas tails, this isophote does not have an elliptical symmetry. To obtain a symmetric isophote, we fit an ellipse to the undisturbed side of the isophote and  replace the isophote on the disturbed side with the ellipse. Everything inside this isophote represents the galaxy disk, the rest constitutes the galaxy ``tail''. {Stellar masses are computed only within the galaxy disk, while for SFR we will also contrast disk and total (disk+tail) values.}

By definition, control sample galaxies have negligible \Ha flux (therefore SFR) in the tails. Quantities for the  stripping sample are given in Tab.\ref{tab:strip}, quantities for the control sample are given in Tab.\ref{tab:contr}.

\section{Results}\label{sec:results}

\begin{table*}[!t]
\caption{Properties of the cluster stripping galaxy sample \label{tab:strip} }
\centering
\begin{tabular}{rrrrrrrr}
\hline
  \multicolumn{1}{c}{ID} &
  \multicolumn{1}{c}{z} &
  \multicolumn{1}{c}{RA} &
  \multicolumn{1}{c}{DEC} &
  \multicolumn{1}{c}{Jstage} &
  \multicolumn{1}{c}{$M_{\ast, disk}$} &
  \multicolumn{1}{c}{$SFR_{disk}$} &
   \multicolumn{1}{c}{$SFR_{tot}$} \\
 \multicolumn{1}{c}{} &
  \multicolumn{1}{c}{} &
  \multicolumn{1}{c}{(J2000)} &
  \multicolumn{1}{c}{(J2000)} &
  \multicolumn{1}{c}{} &
  \multicolumn{1}{c}{($10^{10} M_\odot$)} &
  \multicolumn{1}{c}{($M_\odot \, yr^{-1}$)} &
  \multicolumn{1}{c}{($M_\odot \, yr^{-1}$)} \\
 \hline
  JO10 & 0.0471 & 00:57:41.61 & -01:18:43.994 & 3 & 5.7$\pm$0.8 & 3.1$\pm$0.6 & 3.1$\pm$0.6\\
  JO112 & 0.0583 & 03:40:06.02 & -54:02:27.300 & 1 & 0.41$\pm$0.07 & 0.7$\pm$0.1 & 0.7$\pm$0.1\\
  JO113 & 0.0552 & 03:41:49.17 & -53:24:13.680 & 1 & 0.5$\pm$0.1 & 1.7$\pm$0.3 & 1.8$\pm$0.3\\
  JO13 & 0.0479 & 00:55:39.68 & -00:52:35.981 & 0.5 & 0.7$\pm$0.1 & 1.5$\pm$0.3 & 1.$\pm$0.3\\
  JO135 & 0.0544 & 12:57:04.30 & -30:22:30.313 & 2 & 10$\pm$2 & 4.3$\pm$0.8 & 4.3$\pm$0.9\\
  JO138 & 0.0572 & 12:56:58.51 & -30:06:06.284 & 0.5 & 0.4$\pm$0.1 & 0.22$\pm$0.04 & 0.23$\pm$0.05\\
  JO141 & 0.0587 & 12:58:38.38 & -30:47:32.200 & 1 & 4.8$\pm$1 & 2.5$\pm$0.5 & 2.6$\pm$0.5\\
  JO144 & 0.0515 & 13:24:32.43 & -31:06:59.036 & 1 & 3$\pm$1 & 4.0$\pm$0.8 & 4.0$\pm$0.81\\
  JO147 & 0.0506 & 13:26:49.73 & -31:23:45.511 & 2 & 11$\pm$2 & 4.5$\pm$0.9 & 4.6$\pm$0.9\\
  JO149 & 0.0438 & 13:28:10.53 & -31:09:50.200 & 2 & 0.06$\pm$0.02 & 0.30$\pm$0.06 & 0.42$\pm$0.08\\
  JO156 & 0.0512 & 13:28:34.46 & -31:01:26.777 & 1 & 0.4$\pm$0.1 & 0.28$\pm$0.06 & 0.29$\pm$0.06\\
  JO159 & 0.0480 & 13:26:35.70 & -30:59:36.920 & 1 & 0.7$\pm$0.2 & 1.2$\pm$0.2 & 1.3$\pm$0.3\\
  JO160 & 0.0483 & 13:29:28.62 & -31:39:25.288 & 2 & 1.1$\pm$0.3 & 1.9$\pm$0.4 & 1.9$\pm$0.4\\
  JO162 & 0.0454 & 13:31:29.92 & -33:03:19.576 & 2 & 0.27$\pm$0.06 & 0.40$\pm$0.08 & 0.44$\pm$0.09\\
  JO171 & 0.0521 & 20:10:14.70 & -56:38:30.561 & 2 & 4.1$\pm$0.6 & 1.4$\pm$0.3 & 1.8$\pm$0.4\\
  JO175 & 0.0468 & 20:51:17.60 & -52:49:21.825 & 2 & 3.2$\pm$0.5 & 2.4$\pm$0.5 & 2.5$\pm$0.5\\
  JO179 & 0.0618 & 21:47:07.07 & -43:42:18.221 & 0.5 & 0.33$\pm$0.09 & 0.7$\pm$0.1 & 0.7$\pm$0.1\\
  JO181 & 0.0598 & 22:28:03.80 & -30:18:03.812 & 1 & 0.12$\pm$0.03 & 0.24$\pm$0.05 & 0.26$\pm$0.05\\
  JO194 & 0.0420 & 23:57:00.68 & -34:40:50.117 & 2 & 15.0$\pm$3 & 8$\pm$2 & 9$\pm$2\\
  JO197 & 0.0562 & 09:06:32.58 & -09:31:27.282 & 1 & 1.1$\pm$0.3 & 0.8$\pm$0.2 & 0.8$\pm$0.2\\
  JO200 & 0.0527 & 00:42:05.03 & -09:32:03.841 & 1 & 7$\pm$1 & 2.3$\pm$0.5 & 2.4$\pm$0.5\\
  JO201 & 0.0446 & 00:41:30.29 & -09:15:45.900 & 2 & 6.2$\pm$0.8 & 5$\pm$1 & 6$\pm$1\\
  JO204 & 0.0424 & 10:13:46.83 & -00:54:51.056 & 2 & 4.1$\pm$0.6 & 1.5$\pm$0.3 & 1.7$\pm$0.3\\
  JO206 & 0.0511 & 21:13:47.41 & +02:28:34.383 & 2 & 9.1$\pm$0.9 & 4.8$\pm$0.9 & 5$\pm$1\\
  JO23 & 0.0551 & 01:08:08.10 & -15:30:41.841 & 3 & 0.5$\pm$0.1 & 0.31$\pm$0.06 & 0.31$\pm$0.06\\
  JO27 & 0.0493 & 01:10:48.56 & -15:04:41.611 & 1 & 0.32$\pm$0.08 & 0.5$\pm$0.1 & 0.6$\pm$0.1\\
  JO28 & 0.0543 & 01:10:09.31 & -15:34:24.507 & 1 & 0.23$\pm$0.04 & 0.15$\pm$0.03 & 0.17$\pm$0.03\\
  JO36 & 0.0408 & 01:12:59.42 & +15:35:29.356 & 3 & 6$\pm$1 & 6$\pm$1 & 6$\pm$1\\
  JO47 & 0.0428 & 01:15:57.67 & +00:41:35.938 & 1 & 0.40$\pm$0.06 & 0.39$\pm$0.08 & 0.40$\pm$0.08\\
  JO49 & 0.0451 & 01:14:43.85 & +00:17:10.091 & 1 & 4.8$\pm$0.6 & 1.4$\pm$0.3 & 1.4$\pm$0.3\\
  JO60 & 0.0622 & 14:53:51.57 & +18:39:06.364 & 2 & 2.5$\pm$0.6 & 4.3$\pm$0.9 & 4.5$\pm$0.9\\
  JO69 & 0.0550 & 21:57:19.20 & -07:46:43.794 & 1 & 0.8$\pm$0.2 & 1.3$\pm$0.3 & 1.3$\pm$0.3\\
  JO70 & 0.0578 & 21:56:04.07 & -07:19:38.020 & 1 & 2.9$\pm$0.6 & 2.6$\pm$0.53 & 2.8$\pm$0.5\\
  JO85 & 0.0354 & 23:24:31.36 & +16:52:05.340 & 1 & 4.6$\pm$0.9 & 5$\pm$1 & 6$\pm$1\\
  JO95 & 0.0433 & 23:44:26.66 & +09:06:55.839 & 1 & 0.23$\pm$0.04 & 0.37$\pm$0.07 & 0.40$\pm$0.08\\
  JW10 & 0.0718 & 04:39:18.19 & -21:57:49.627 & 1 & 1.0$\pm$0.2 & 0.8$\pm$0.1 & 0.8$\pm$0.2\\
  JW100 & 0.0619 & 23:36:25.06 & +21:09:02.529 & 2 & 29$\pm$7 & 2.6$\pm$0.5 & 4.$\pm$0.8\\
  JW108 & 0.0477 & 06:00:47.96 & -39:55:07.416 & 3 & 3.0$\pm$0.7 & 2.2$\pm$0.4 & 2.2$\pm$0.4\\
  JW115 & 0.0725 & 12:00:47.95 & -31:13:41.635 & 1 & 0.5$\pm$0.1 & 0.5$\pm$0.1 & 0.5$\pm$0.1\\
  JW29 & 0.0431 & 12:57:49.48 & -17:39:57.095 & 0.5 & 0.32$\pm$0.06 & 0.5$\pm$0.1 & 0.5$\pm$0.1\\
  JW39 & 0.0663 & 13:04:07.71 & +19:12:38.486 & 2 & 17$\pm$3 & 3.1$\pm$0.6 & 3.6$\pm$0.7\\
  JW56 & 0.0387& 13:27:03.03 & -27:12:58.205 & 2 & 0.11$\pm$0.03 & 0.13$\pm$0.03 & 0.17$\pm$0.03\\
 \hline
\end{tabular}
\end{table*}

\begin{table*}
\caption{Properties of the control galaxy sample \label{tab:contr}}
\centering
\begin{tabular}{rrrrrrr}
\hline
  \multicolumn{1}{c}{ID} &
  \multicolumn{1}{c}{z} &
  \multicolumn{1}{c}{RA} &
  \multicolumn{1}{c}{DEC} &
  \multicolumn{1}{c}{$M_{\ast, disk}$} &
  \multicolumn{1}{c}{$SFR_{disk}$} &
  \multicolumn{1}{c}{$SFR_{tot}$} \\
  \multicolumn{1}{c}{} &
  \multicolumn{1}{c}{} &
  \multicolumn{1}{c}{(J2000)} &
  \multicolumn{1}{c}{(J2000)} &
  \multicolumn{1}{c}{($10^{10} M_\odot$)} &
  \multicolumn{1}{c}{($M_\odot \, yr^{-1}$)} &
  \multicolumn{1}{c}{($M_\odot \, yr^{-1}$)} \\
 \hline
  \multicolumn{7}{c}{cluster}  \\
 A3128\_B\_0148 & 0.0575 & 03:27:31.09 & -52:59:07.655 & 0.7$\pm$0.2 & 0.7$\pm$0.1 & 0.7$\pm$0.14\\
  A3266\_B\_0257 & 0.0584 & 04:27:52.58 & -60:54:11.565 & 0.8$\pm$0.2 & 0.5$\pm$0.1 & 0.5$\pm$0.1\\
  A3376\_B\_0261 & 0.0506 & 06:00:13.68 & -39:34:49.232 & 3.4$\pm$0.6 & 2.0$\pm$0.4 & 2.1$\pm$0.4\\
  A970\_B\_0338 & 0.0591 & 10:19:01.65 & -10:10:36.924 & 1.2$\pm$0.2 & 0.7$\pm$0.1 & 0.7$\pm$0.1\\
  JO102 & 0.0594 & 03:29:04.69 & -52:50:05.364 & 1.0$\pm$0.2 & 0.9$\pm$0.2 & 0.9$\pm$0.2\\
  JO123 & 0.0550 & 12:53:01.03 & -28:36:52.584 & 0.7$\pm$0.2 & 0.25$\pm$0.05 & 0.25$\pm$0.05\\
  JO128 & 0.0500 & 12:54:56.84 & -29:50:11.184 & 0.8$\pm$0.2 & 0.8$\pm$0.2 & 0.8$\pm$0.2\\
  JO17 & 0.0451 & 01:08:35.33 & +01:56:37.043 & 1.4$\pm$0.3 & 0.9$\pm$0.2 & 0.9$\pm$0.2\\
  JO180 & 0.0647 & 21:45:15.00 & -44:00:31.188 & 1.0$\pm$0.2 & 0.42$\pm$0.08 & 0.42$\pm$0.08\\
  JO205 & 0.0448 & 21:13:46.12 & +02:14:20.355 & 0.33$\pm$0.08 & 0.6$\pm$0.1 & 0.6$\pm$0.1\\
  JO41 & 0.0477 & 12:53:54.79 & -15:47:20.096 & 1.6$\pm$0.3 & 0.31$\pm$0.06 & 0.31$\pm$0.06\\
  JO45 & 0.0425 & 01:13:16.58 & +00:12:05.839 & 0.15$\pm$0.04 & 0.11$\pm$0.02 & 0.12$\pm$0.02\\
  JO5 & 0.0648 & 10:41:20.38 & -08:53:45.559 & 1.9$\pm$0.4 & 1.3$\pm$0.3 & 1.3$\pm$0.3\\
  JO68 & 0.0561 & 21:56:22.00 & -07:54:28.971 & 1.0$\pm$0.2 & 0.7$\pm$0.1 & 0.7$\pm$0.1\\
  JO73 & 0.0713 & 22:04:25.99 & -05:14:47.041 & 1.1$\pm$0.3 & 1.0$\pm$0.2 & 1.0$\pm$0.2\\
  JO89 & 0.0423 & 23:26:00.60 & +14:18:26.291 & 0.5$\pm$0.1 & 0.12$\pm$0.02 & 0.12$\pm$0.02\\
  JO93 & 0.0370 & 23:23:11.74 & +14:54:05.013 & 3.5$\pm$0.5 & 1.8$\pm$0.4& 1.9$\pm$0.4\\
  \multicolumn{7}{c}{field}  \\
  P13384 & 0.0512 & 10:53:03.15 & -00:13:30.932& 0.7$\pm$0.2 & 0.6$\pm$0.1 &  0.6$\pm$0.1\\
  P15703 & 0.0424 & 11:06:33.28 & +00:16:48.192& 10$\pm$2    & 1.1$\pm$0.2     & 1.1$\pm$0.2\\
  P17945 & 0.0439 & 11:15:26.45 & +00:16:11.586 & 0.6$\pm$0.1 & 0.49$\pm$0.09  & 0.5$\pm$0.1\\
  P19482 & 0.0407 & 11:22:31.25 & -00:01:01.601 & 2.2$\pm$0.4 & 1.3$\pm$0.2    & 1.3$\pm$0.2\\
  P20769 & 0.0489 & 11:27:17.60 & +00:11:24.388 & 0.3$\pm$0.1 & 0.19$\pm$0.04  & 0.20$\pm$0.04\\
  P20883 & 0.0614 & 11:27:45.41 & -00:07:16.580 & 0.8$\pm$0.2 & 0.37$\pm$0.07  & 0.37$\pm$0.07\\
  P21734 & 0.0686 & 11:31:07.90 & -00:08:07.914 & 6$\pm$1     & 3.0$\pm$0.6     & 3.1$\pm$0.6\\
  P25500 & 0.0603 & 11:51:36.28 & +00:00:01.929 & 7$\pm$1     & 2.1$\pm$0.4      & 2.2$\pm$0.4\\
  P42932 & 0.0410 & 13:10:44.71 & +00:01:55.540 &3.3$\pm$0.6 & 2.2$\pm$0.4   & 2.2$\pm$0.4\\
  P45479 & 0.0515 & 13:23:34.73 & -00:07:51.673 & 3.7$\pm$0.6 & 1.8$\pm$0.4    & 1.8$\pm$0.4\\
  P48157 & 0.0614 & 13:36:01.59 & +00:15:44.696 & 3.9$\pm$0.8 & 2.5$\pm$0.5    & 2.5$\pm$0.5\\
  P57486 & 0.0529 & 14:11:34.45 & +00:09:58.293 &0.9$\pm$0.2 & 0.6$\pm$0.1  & 0.6 $\pm$0.1\\
  P648 & 0.0661 & 10:01:27.74 & +00:09:18.372 &  2.8$\pm$0.6 & 1.1$\pm$0.2   & 1.1$\pm$0.2\\
  P669 & 0.0457 & 10:02:00.62 & +00:10:44.299 & 3.2$\pm$0.6 & 0.7$\pm$0.1   & 0.7$\pm$0.1\\
  P954 & 0.0451 & 10:02:03.33 & -00:12:49.836 & 0.4$\pm$0.1 & 0.38$\pm$0.07  & 0.38$\pm$0.07\\
\hline\end{tabular}
\end{table*}

\begin{figure}
\centering
\includegraphics[scale=0.35]{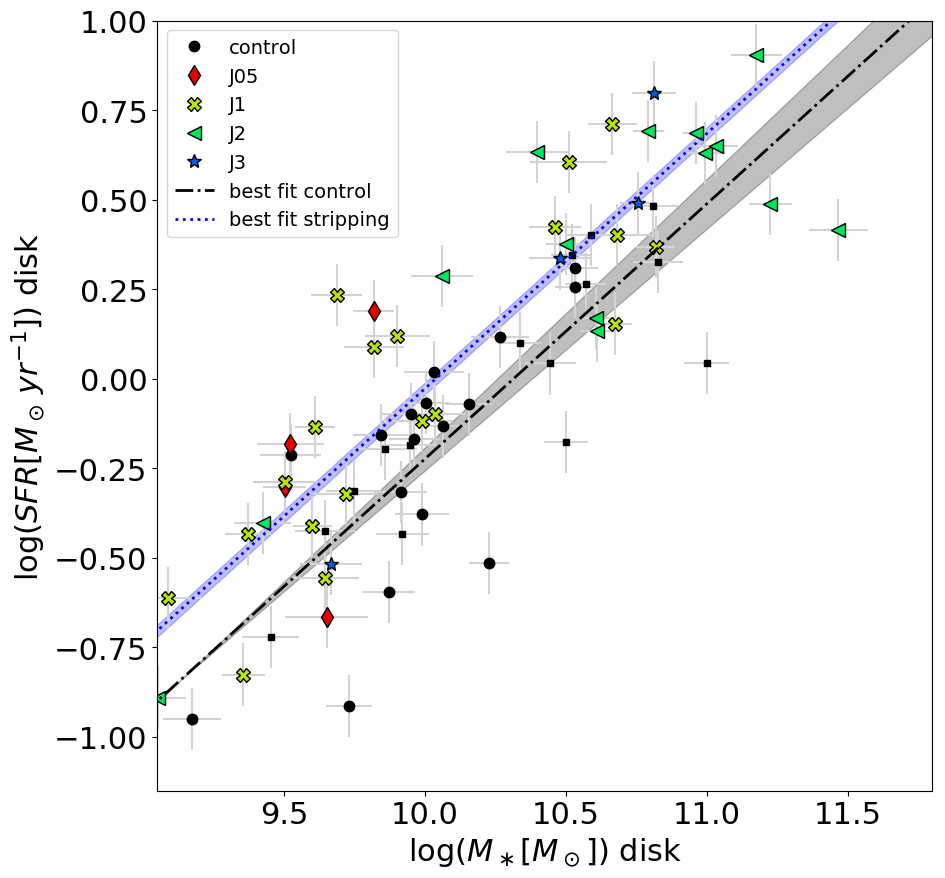}
\caption{ \label{fig:SFR_mass} Disk SFR-Mass relation for stripping and control sample galaxies. Black symbols refer to the control sample, with dots representing galaxies in  clusters, squares galaxies in the field. Colored symbols represent stripping galaxies, with  red diamonds representing  galaxies at initial stripping (J0.5), yellow crosses stripping galaxies  (J1), green triangles peak stripping galaxies (J2), light blue stars truncated disks (J3). Black dash dotted line and shaded gray area show the best fit for the control sample, blue dashed line and shaded blue area show the best fit for the stripping sample, adopting the same slope as the control sample.}
\end{figure}

The main result of the paper is shown in Fig.\ref{fig:SFR_mass}, where the disk SFR-M$_\ast$ relation for galaxies at different stages of stripping is shown. These galaxies are compared to the control sample. 

In the control sample, field and cluster galaxies occupy the same region of the plot, indicating that undisturbed star forming galaxies do not feel the environment in which they are embedded yet and that cluster galaxies can be as star forming as field galaxies, in agreement with many literature results \citep[e.g.,][]{Paccagnella2016}. 
The relation is qualitatively in agreement with the one shown in \cite{Poggianti2016}, even though a meaningful quantitative comparison is prevented by the differences on the data acquisition and analysis, as discussed in the Introduction (see Boselli et al. 2015, Richards et al. 2015, Gavazzi et al. 2018 for a discussion on the comparison between nuclear and total spectra). 

It seems immediately clear, instead, that stripping galaxies  populate the upper envelope of the control sample relation. 
J1 and J2 galaxies, that are the objects with the longest tails, deviate the most. 
Nonetheless, even truncated disks (J3) and galaxies at initial stripping (J0.5) do show a significant enhancement of the SF. 
A mass segregation effect is also visible: J2 galaxies are also the most massive ones. 

{Results are not driven by the galaxy inclination angles: no significant trends are observed when inclination is taken into account (plot not shown).}

To place the differences between the two samples on a statistical ground, we fit the datapoints to obtain the best fit relations, using a least square fitting method that takes into account uncertainties on both  axis. When fitting the control sample, we assume both  parameters free, when instead we fit the stripping sample we use the slope of the control sample, to allow a fairer comparison of the intercepts. We  do not attempt to fit separately galaxies belonging to different  classes, due to the low number statistics. 

The control  and stripping samples are described by relations that are different at more than 2$\sigma$ levels, and the difference between the two fits is {0.2} dex. 

\begin{figure}
\centering
\includegraphics[scale=0.35]{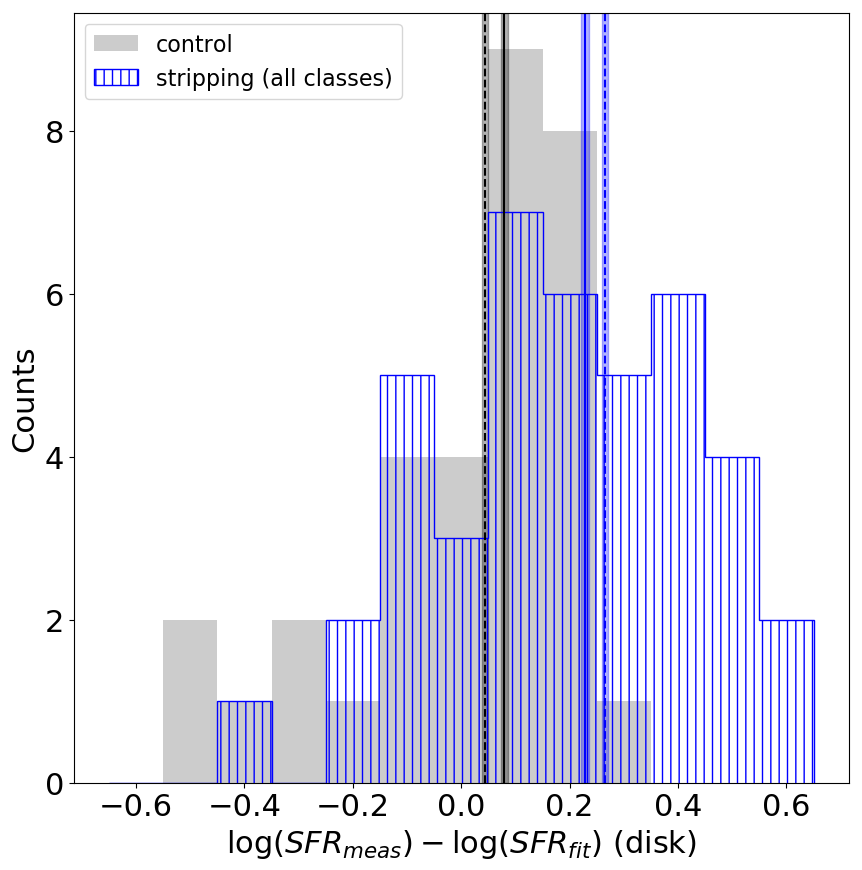}
\caption{Distributions of the differences between the galaxy SFRs and their expected value according to the fit to the control sample, given their mass. Black and gray colors refer to the control sample, blue color refers to the stripping sample. Dashed lines give mean values, solid lines median values.   \label{fig:SFR_diff} }
\end{figure}

This result indicates that galaxies feeling the effect of RPS show a significant enhancement of the SFR on the disk of the galaxy, with respect to control sample galaxies of similar mass.  Differences are better seen in Fig. \ref{fig:SFR_diff}, where the distribution of the difference between the  SFR of each galaxy and the value derived from the control sample fit given the galaxy mass is shown. As the fit was carried out taking into account the  uncertainties in both $M_\ast$ and SFR, the control sample distribution does not peak exactly at zero.
A tail of galaxies with reduced SFR is visible in the control sample. Two of these galaxies belong to the field, two to the cluster sample. Even though their current SF map does not show anomalies, their spatially resolved SF histories obtained from \sinopsis show that the radial extent of the SF has reduced in the recent past, suggesting that these galaxies are transitioning from being star forming to passive \citep{Paccagnella2016}. 
Removing these galaxies reduces the differences between stripping and control galaxies discussed in Fig.\ref{fig:SFR_mass}, but fits are still different at $>$1$\sigma$ level and  their difference is  $\sim 0.1$ dex. 

In contrast, most of the stripping galaxies have a measured SFR higher than that expected given the fit, with the results that their distribution is skewed towards higher values and is also broader. The standard deviation of the stripping sample is indeed $\sim$1, while that of the control sample is 0.4. The Kolmogorov-Smirnov test gives a high probability distributions are different (pvalue$<$0.0001).

Both median and mean values are statistically different: $\Delta(median)=0.15\pm0.01$, $\Delta(mean)=0.22\pm0.01$. Note that the mean values are even more different because they are influenced by the tail of objects in the control sample with $\log(SFR_
{meas})-\log(SFR_{fit})\sim -0.4$.

Having assessed that the SFR is enhanced in the disk, we can also investigate how much SF  is found in the tails, for the stripping galaxies. We remind the reader that in the control sample the amount of SF  outside the disk is negligible and galaxies are located approximately on the 1:1 relation.

\begin{figure}
\centering
\includegraphics[scale=0.35]{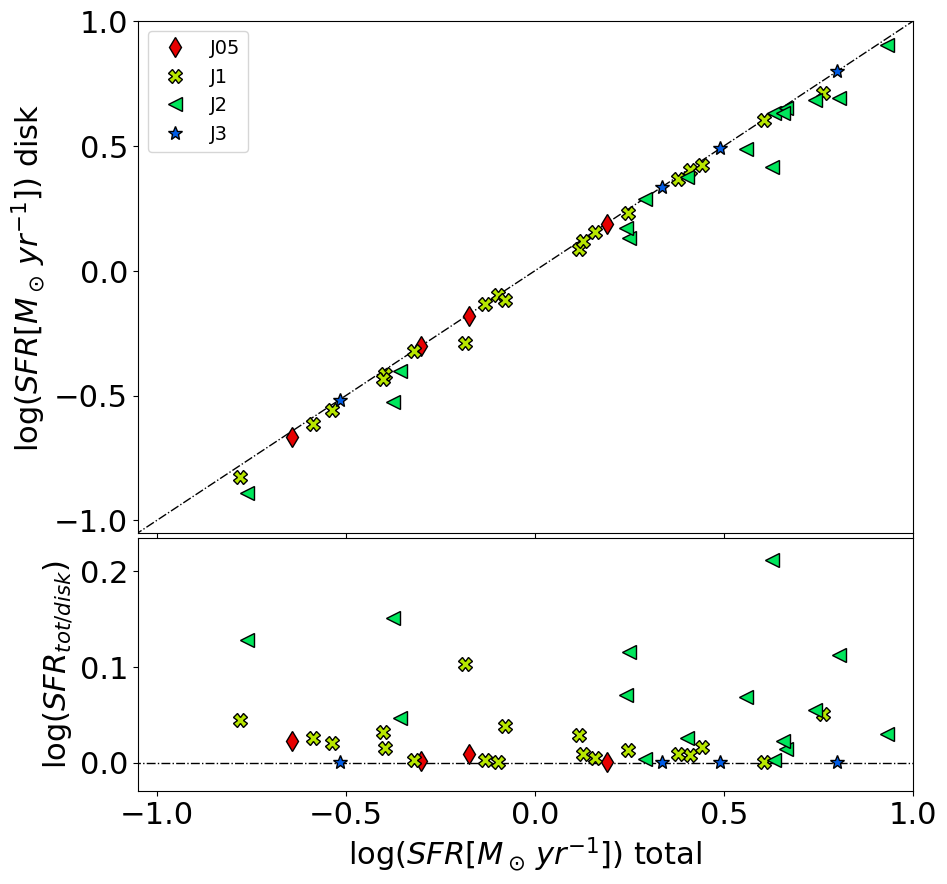}
\caption{Comparison between the total and disk SFR. The upper panel compares the actual values, the lower panel shows the difference as a function of the total SFR. Dotted lines represent the loci where total and disk values coincide. Colors and symbols are as in Fig.\ref{fig:SFR_mass}.  \label{fig:cfr_values} }
\end{figure}

Figure \ref{fig:cfr_values} compares the total and disk SFR values 
of the stripping galaxies. 
As expected, in these galaxies the contribution of the SFR outside the disk is conspicuous. 
The ratio between the total and disk SFR does not depend on the total SFR, indicating that the fraction of new stars produced in the tails is not strictly related to the total  SFR. We do not detect a trend of  the total and disk SFR with mass either (plot not shown).  
A dependence on the Jstage is instead observed: galaxies at the initial stripping and truncated disk have SFR$_{tot}$$\sim$SFR$_{disk}$, indicating no activity outside the disk. In contrast, in galaxies classified as J1 and J2, up to 1/3 of the total SFR can occur in the wake.  Considering all classes together, the median $\log$ SFR difference between the two components is 0.02.

The differences between total and disk values are  overall quite small, therefore the SFR-M$_\ast$ relation obtained using the total values is not significantly different from the one presented in Fig.\ref{fig:SFR_mass} (plot not shown). 

To conclude, our results show that in stripping galaxies, at any given mass, the SFR  enhancement is already significant taking into account only the SF in the disk. 

\section{Summary and Conclusions}\label{sec:dc}

Exploiting the GASP dataset, we have shown for the first time {in a statistically significant sample,} that in  observations stripping galaxies are  characterised by a systematic enhancement of the SFR, compared to undisturbed galaxies of similar mass, indicating that SF is boosted in the disks during stripping. Additional SF takes place in the tails. {We therefore confirm the results of \cite{Poggianti2016}. Such enhancement had been already observed, but only in few individual cases \citep[e.g.][]{Crowl2006,Merluzzi2016,Kenney2014}. }

Due to otherwise low sample statistics, our control sample includes both cluster and field galaxies. It is still controversial whether the SFR-M$_\ast$ relation is dependent or not on environment and by how much. Results of e.g. \cite{Paccagnella2016, Vulcani2010} showed that in  cluster cores the relation is systematically shifted towards lower values, due to a population of galaxies with suppressed SF. Nonetheless, the bulk of the cluster population can be as star forming as field galaxies \citep[see also][]{Poggianti2006}. If, overall, the cluster relation is shifted low with respect to field galaxies, ours can be seen as a lower limit of the gap between stripping and undisturbed galaxies in clusters, and differences might be even more striking. 

Our results suggest that RP moderately enhances the SF also in the central disk before removing the gas.

The triggering of SF in stripping galaxies has been debated in simulations, with different authors reaching quite different conclusions, as summarised in the Introduction. The origin of the enhancement is not  clear yet, with some theoretical studies predicting a correlation with the cluster dynamical state. In future papers we will investigate the conditions in which SF triggering happens (relaxed or dynamically disturbed environment).

In Gullieuszik et al. (in prep.) we will analyse whether the  amount of mass and SF in the tails  depends also on other factors, such as the cluster velocity dispersion, the position of the galaxy within the cluster, its orbit and its relative velocity with respect to the hosting system.

\acknowledgments
We thank the referee for their useful comments. Based on observations collected at the European Organisation for Astronomical Research in the Southern Hemisphere under ESO programme 196.B-0578. We acknowledge funding from the INAF PRIN-SKA 2017 program 1.05.01.88.04. 
Y.~J. acknowledges support from CONICYT PAI (Concurso Nacional de Inserci\'on en la Academia 2017) No. 79170132.

\bibliography{gasp}

\end{document}